\documentclass[12pt,onecolumn,amsmath,amssymb]{revtex4}
\usepackage{graphicx}
\begin{document}

\title{Formation of Conic Cusps at the Surface of Liquid Metal
in Electric Field}

\author{Nikolay M. Zubarev}
\email{nick@ami.uran.ru}

\affiliation{Institute of Electrophysics, Ural Branch, Russian
Academy of Sciences,\\ 106 Amundsen Street, 620016 Ekaterinburg, Russia}

\begin{abstract}

The formation dynamics is studied for a singular profile of a surface of
an ideal conducting fluid in an electric
field. Self-similar solutions of electrohydrodynamic equations describing
the fundamental process of formation
of surface conic cusps with angles close to the Taylor cone angle
$98.6^{\circ}$ are obtained. The behavior of physical
quantities (field strength, fluid velocity, surface curvature) near
the singularity is established.

\end{abstract}

\maketitle

It is known \cite{1, 2} that a flat boundary of liquid metal
becomes unstable in a strong electric field. The development
of instability results in conic cusp singularities,
from which the strengthened field initiates emission
processes \cite{3,4,5,6}. The description of these processes is a
key problem of the electrohydrodynamics of conducting
fluids with free surfaces; interest in this problem is
largely caused by the practical use of liquid-metal
sources of charged particles. The progress in this field
is associated with Taylor's work \cite{7}, where it was demonstrated
that the surface electrostatic pressure
$P_E$ for a cone with angle $98.6^{\circ}$
depends on the distance from its
axis as $r^{-1}$
and, hence, can be counterbalanced by the
surface pressure $P_S\sim r^{-1}$.
Since the force balance is violated
at the cone apex, Taylor's solution cannot be
treated as the exact solution of the problem of equilibrium
configuration of a charged surface of conducting
fluid and only represents the possible asymptotic form
at $r\to\infty$.
At the same time, it turned out that Taylor's
solution nicely describes the experimentally observed
surface shape before the instant of singularity formation.
It was pointed out in \cite{3,4,5,6} that the angle of incipient
conic formations is close to the Taylor cone angle.

What is the reason for such a coincidence? One may
assume that the mechanism for the formation of conic
cusps with an angle of $98.6^{\circ}$ during a finite time is not
directly associated with the static Taylor model. A high
reproducibility of experimental results and a weak
dependence of fluid behavior at the final instability
stages on the geometry of the system suggest that the
behavior of fluid near the singularity has a self-similar
character.

Let us check the validity of this hypothesis. Consider
the potential motion of an ideal fluid occupying
the region bounded by free surface
$z=\eta(x,y,t)$.
We will assume that the vector of an external electric field is
directed along the $z$ axis and equals $E$. The velocity
potential $\Phi$ of fluid and the electric-field potential $\varphi$
satisfy the Laplace equations
$$
\nabla^2\Phi=0, \qquad \nabla^2\varphi=0
$$
with the following boundary conditions:
$$
\Phi_t+\frac{|\nabla\Phi|^2}{2}=
\frac{|\nabla\varphi|^2}{8\pi\rho}+
\frac{\alpha}{\rho}\,\nabla_{\!\!\bot}\cdot
\frac{\nabla_{\!\!\bot}\eta}{\sqrt{1+(\nabla_{\!\!\bot}\eta)^2}},
\qquad z=\eta(x,y,t),
$$
$$
\eta_t=\Phi_z-\nabla_{\!\!\bot}\eta\cdot\nabla_{\!\!\bot}\Phi,
\qquad z=\eta(x,y,t),
$$
$$
\varphi=0, \qquad z=\eta(x,y,t),
$$
$$
|\nabla\Phi|\to 0, \qquad z\to-\infty,
$$
\begin{equation}
\varphi\to -Ez, \qquad z\to\infty,
\end{equation}
where $\alpha$ is the surface tension coefficient and $\rho$
is the density of a medium.

We are interested in the dynamics of formation of a
singular profile for a conducting fluid. It is natural to
assume that the electric field near the cusp appreciably
exceeds the external field; i.e., $|\nabla\varphi|\gg E$.
In this case,
the interface evolution is fully determined by the intrinsic
field, which decreases with distance from the singularity.
One can thus use the condition
\begin{equation}
|\nabla\varphi|\to 0, \qquad z\to\infty
\end{equation}
instead of the field uniformity condition (1). This
agrees with the assumption about the universal behavior
of a fluid in the formation of a singular surface
profile, because it allows the fluid motion near the singular
point to be analyzed without regard for the particular
geometry of the problem. The applicability of condition
(2) will be discussed below in more detail after
establishing some regularities for the dynamics of a
conducting fluid near the singularity. Note that the possibility
of secondary Taylor cones nucleating at the
already formed cones counts in favor of the universal
mechanism of formation of conic cusps \cite{4}. It is clear
from this example that fluid "forgets" the boundary
conditions at infinity at the stage of collapse.

Let us consider the most important case of the axially
symmetric perturbation of the surface. Taking into
account that, after substitutions
$$
\varphi\to\varphi\,4\pi\alpha E^{-1}, \qquad
\Phi\to\Phi\,2\pi^{\frac{1}{2}}\rho^{-\frac{1}{2}}\alpha E^{-1}, \qquad
\eta\to\eta\,4\pi\alpha E^{-2},
$$
$$
r\to r\,4\pi\alpha E^{-2}, \qquad
z\to z\,4\pi\alpha E^{-2}, \qquad
t\to t\,8\pi^{\frac{3}{2}}\rho^{\frac{1}{2}}\alpha E^{-3},
$$
where $r=\sqrt{x^2+y^2}$,
the equations of motion become
dimensionless and do not contain any physical characteristics,
one obtains
\begin{equation}
\Phi_{rr}+r^{-1}\Phi_r+\Phi_{zz}=0, \qquad z<\eta(r,t),
\end{equation}
\begin{equation}
\varphi_{rr}+r^{-1}\varphi_r+\varphi_{zz}=0, \qquad z>\eta(r,t),
\end{equation}
\begin{equation}
\Phi_t+\frac{\Phi_r^2+\Phi_z^2}{2}=
\frac{\varphi_r^2+\varphi_z^2}{2}+
\frac{1}{\sqrt{1+{\eta_r}^2}}\,
\left(\frac{\eta_{rr}}{1+{\eta_r}^2}+\frac{\eta_r}{r}\right),
\qquad z=\eta(r,t),
\end{equation}
\begin{equation}
\eta_t=\Phi_z-\eta_r\Phi_r, \qquad z=\eta(r,t),
\end{equation}
\begin{equation}
\varphi=0, \qquad z=\eta(r,t),
\end{equation}
\begin{equation}
\Phi_r^2+\Phi_z^2\to 0, \qquad
\varphi_r^2+\varphi_z^2\to 0, \qquad r^2+z^2\to\infty,
\end{equation}
\begin{equation}
\Phi_r=0, \qquad \varphi_r=0, \qquad \eta_r=0, \qquad r=0.
\end{equation}
These equations allow the only self-similar substitution
\begin{equation}
\Phi(x,y,z,t)=\tilde{\Phi}(\tilde{r},\tilde{z})\tau^{1/3},
\end{equation}
\begin{equation}
\varphi(x,y,z,t)=\tilde{\varphi}(\tilde{r},\tilde{z})\tau^{1/3},
\end{equation}
\begin{equation}
\eta(x,y,t)=\tilde{\eta}(\tilde{r})\tau^{2/3},
\end{equation}
\begin{equation}
r=\tilde{r}\tau^{2/3},
\end{equation}
\begin{equation}
z=\tilde{z}\tau^{2/3},
\end{equation}
where $\tau=t_c-t$ and $t_c$ is the collapse time. This substitution
occurs due to the fact that Eqs. (3)--(9) are invariant
about dilatations
$$
\Phi\to\Phi c, \qquad
\varphi\to\varphi c, \qquad
\eta\to\eta c^2,
$$
$$
r\to r c^2, \qquad
z\to z c^2, \qquad
t\to t c^3,
$$
i.e., it occurs, in fact, from dimensional considerations
(c is an arbitrary constant). Note that the initial electrohydrodynamic
equations with condition (1) do not permit
one to introduce any self-similar variables.

Substituting Eqs. (10)--(14) in Eqs. (3)--(9), one
finds that the functions 
$\tilde{\Phi}$, $\tilde{\varphi}$ and $\tilde{\eta}$ 
obey the following set of partial differential equations:
\begin{equation}
\tilde{\Phi}_{\tilde{r}\tilde{r}}+
\tilde{r}^{-1}\tilde{\Phi}_{\tilde{r}}
+\tilde{\Phi}_{\tilde{z}\tilde{z}}=0, \qquad
\tilde{z}<\tilde{\eta}(\tilde{r}),
\end{equation}
\begin{equation}
\tilde{\varphi}_{\tilde{r}\tilde{r}}+
\tilde{r}^{-1}\tilde{\varphi}_{\tilde{r}}
+\tilde{\varphi}_{\tilde{z}\tilde{z}}=0, \qquad
\tilde{z}>\tilde{\eta}(\tilde{r}),
\end{equation}
\begin{equation}
\frac{2\tilde{\Phi}_{\tilde{r}}\tilde{r}+
2\tilde{\Phi}_{\tilde{z}}\tilde{\eta}-\tilde{\Phi}}{3}
+\frac{\tilde{\Phi}_{\tilde{r}}^2+\tilde{\Phi}_{\tilde{z}}^2}{2}
=\frac{\tilde{\varphi}_{\tilde{r}}^2+
\tilde{\varphi}_{\tilde{z}}^2}{2}+
\frac{1}{\sqrt{1+\tilde{\eta}_{\tilde{r}}^2}}\,
\left(\frac{\tilde{\eta}_{\tilde{r}\tilde{r}}}
{1+\tilde{\eta}_{\tilde{r}}^2}+\frac{\tilde{\eta}_{\tilde{r}}}{\tilde{r}}
\right),
\qquad \tilde{z}=\tilde{\eta}(\tilde{r}),
\end{equation}
\begin{equation}
2\tilde{\eta}_{\tilde{r}}\tilde{r}-2\tilde{\eta}=
3\tilde{\Phi}_{\tilde{z}}-
3\tilde{\eta}_{\tilde{r}}\tilde{\Phi}_{\tilde{r}},
\qquad \tilde{z}=\tilde{\eta}(\tilde{r}),
\end{equation}
\begin{equation}
\tilde{\varphi}=0, \qquad \tilde{z}=\tilde{\eta}(\tilde{r}),
\end{equation}
\begin{equation}
\tilde{\Phi}_{\tilde{r}}^2+\tilde{\Phi}_{\tilde{z}}^2\to 0, \qquad
\tilde{\varphi}_{\tilde{r}}^2+\tilde{\varphi}_{\tilde{z}}^2\to 0, \qquad 
\tilde{r}^2+\tilde{z}^2\to\infty,
\end{equation}
\begin{equation}
\tilde{\Phi}_{\tilde{r}}=0, \qquad \tilde{\varphi}_{\tilde{r}}=0, 
\qquad \tilde{\eta}_{\tilde{r}}=0, \qquad \tilde{r}=0.
\end{equation}
For self-similar solutions (10)--(14), the surface profile
forms first at the periphery and then extends to the center
$r=z=0$ (the scale decreases as $\tau^{2/3}$). 
This implies that the formation of conic cusps at $t=t_c$ 
is described
by those solutions to the set of Eqs. (15)--(21) which
provide conic asymptotic shape of the surface. In such
a situation, the presence of asymptotic solutions for
which $\tilde{\eta}\sim\tilde{r}$ at $\tilde{r}\to\infty$
is the necessary condition for
the validity of our assumption about the self-similar
nature of conic formations.

Analysis of Eqs. (15)--(21) in the limit 
$R=\sqrt{\tilde{r}^2+\tilde{z}^2}\to\infty$
showed that they have an asymptotic solution of the form
\begin{equation}
\tilde{\varphi}(\tilde{r},\tilde{z})=p^{-1}\!
\left[2R(s_0-s)\right]^{1/2}\!P_{1/2}(\cos\theta),
\end{equation}
\begin{equation}
\tilde{\Phi}(\tilde{r},\tilde{z})=sR^{-1}, 
\end{equation}
\begin{equation}
\tilde{\eta}(\tilde{r})=-s_0\tilde{r},
\end{equation}
$$
P_{1/2}(\cos\theta_0)=0,
$$
$$
p=\left[dP_{1/2}(\cos\theta)/d\theta\right]_{\theta=\theta_0},
$$
$$
s_0=-\mbox{ctg}\,\theta_0,
$$
where $\theta=\mbox{arctg}\,(\tilde{r}/\tilde{z})$ 
is the polar distance in spherical coordinates, 
$P_{1/2}$ is the Legendre polynomial of order
$\frac{1}{2}$, and $s$ is a constant satisfying inequality $0<s<s_0$.
This solution describes a conic surface with an angle of
$2\pi-2\theta_0$ that is equal to approximately $98.6^{\circ}$, i.e., to the
Taylor cone angle. According to Eq. (23), the fluid
motion is spherically symmetric, and fluid moves to the
sink point $R=0$ along the tangent to the surface (24).
Since the self-similar solution assumes its asymptotic
form at $\tau\to 0$, a conic cusp with Taylor angle forms
at time $t_c$, and Eqs. (22)--(24) are the exact analytic
solution of the problem. The electric field at the cusp
increases as $\tau^{-1/3}$, the cusp growth velocity increases as
$\tau^{-1/3}$, and the cusp curvature increases as $\tau^{-2/3}$;
i.e., these quantities become infinite during a finite time. At an
appreciable distance from the singularity, the field
strength does not change, and the velocity of fluid linearly
decreases with time and becomes zero at 
$t=t_c$. The latter fact allows one to explain qualitatively the
mechanism of transition to the stationary regime occurring
for liquid-metal sources after the initiation of field
ion evaporation from the cusp (stationary emitter models
were developed in \cite{8, 9}).

This analysis is only valid on the condition that the
asymptotic solutions to the set of partial differential
Eqs. (15)--(21) have the form of Eqs. (22)--(24). To
prove this statement, one should construct the asymptotic
expansion for the solutions at $R\to\infty$
with leading terms given by Eqs. (22)--(24). Let us seek this
expansion in the form
\begin{equation}
\tilde{\varphi}(\tilde{r},\tilde{z})=\sum_{n=0}^{\infty}
a_n\,\frac{\partial^{3n}}{\partial {\tilde{z}}^{3n}}
\left[R^{1/2}P_{1/2}(\cos\theta)\right],
\end{equation}
\begin{equation}
\tilde{\Phi}(\tilde{r},\tilde{z})=\sum_{n=0}^{\infty}
b_n\,\frac{\partial^{3n}}{\partial {\tilde{z}}^{3n}}
\left[R^{-1}\right], 
\end{equation}
\begin{equation}
\tilde{\eta}(\tilde{r})=\sum_{n=0}^{\infty}
c_n\,{\tilde{r}}^{1-3n},
\end{equation}
where it is taken into account that the derivative of a
harmonic function of any order with respect to $\tilde{z}$ is also
a harmonic function. The zero-order coefficients are
determined by Eqs. (22)--(24),
$$
a_0=p^{-1}\!\left[2(s_0-s)\right]^{1/2}, \qquad b_0=s, \qquad c_0=s_0.
$$
It turns out that, to the first order, the surface is conic:
$$
a_1=0, \qquad 
b_1=-\frac{s^2(1+s_0^2)^{3/2}}{18s_0(3-2s_0^2)}, 
\qquad c_1=0.
$$
The correction to Eq. (24) for the surface shape appears
in the next order. One finds from kinematic boundary
condition (18) that
$$
c_2=-\frac{s^2(4s_0^2-1)}{8s_0^2\left(1+s_0^2\right)^2(3-2s_0^2)}.
$$
The coefficients  $a_2$ and $b_2$ can be determined from
Eq. (19) and, correspondingly, Eq. (17), where one
should use the linear order of perturbation theory for
small deviation of the surface from the cone. In turn,
the coefficient $c_3$ is found from Eq. (18), where one
should take into account the quadratic nonlinearity, etc.
Thus, the expansion coefficients are uniquely determined
by the zero-order coefficients, i.e., by the parameter
$s$ of the problem, confirming the existence of solutions
with the desired asymptotic form for the equations
of motion. Note that, if one sets $s=0$ and, hence,
$\Phi=0$, then Eqs. (15)--(21) coincide with the Taylor
equations in the problem of equilibrium configuration
of a charged liquid-metal surface. However, one fails to
construct asymptotic expansion (25)--(27) in this case.

\begin{figure}
\includegraphics{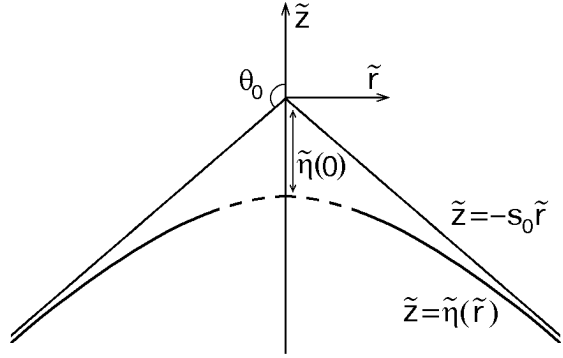}
\caption{\label{fig:figure}
Schematic drawing of a surface of conducting fluid  
$\tilde{z}=\tilde{\eta}(\tilde{r})$
corresponding to the self-similar solutions of the
equations of motion.
}
\end{figure}

Further, the solutions given by expansions (25)-(27)
adequately describe the experimental data if the condition 
$\tilde{\eta}<-s_0\tilde{r}$ is satisfied, 
i.e., if the fluid surface is
positioned above the asymptotic cone (see figure~\ref{fig:figure}).
Otherwise, the surface velocity would be directed in opposition
to the $z$ axis. Let us check how this condition is
fulfilled in the limit of large $r$. It follows from the
expansion obtained above for the surface shape that it
deviates from the conic shape in the direction specified
by the sign of the $c_2$ coefficient. In our case, 
$c_2<0$ ($c_2>0$ for cone angles smaller than $78.5^{\circ}$ 
and larger than $126.9^{\circ}$),
so that the amplitude of surface perturbation
should increase. Indeed, the evolution of the fluid
boundary away from the singularity is determined by
the leading terms of the expansion in small $\tau$ 
value,
$$
\eta(r,t)=-s_0r-|c_2|\tau^4r^{-5},
$$
from whence it follows that, when forming a conic
cusp, the fluid moves upwards, as is expected from
physical considerations.

Let us now consider the surface geometry for small $\tilde{r}$,
where expansion (27) diverges. For the function $\tilde\eta$
to satisfy condition (21), the surface near the cone apex
must be "rounded off" (figure). Let us estimate the distance
$|\tilde\eta(0)|$ from the cone apex to the fluid surface.
Multiplying kinematic boundary condition (18) by
$2\pi\tilde{r}/3$ and integrating it over $\tilde r$,
one obtains after simple
mathematics
$$
2V=\int_S\partial_n\tilde{\Phi}\,dS,
$$
where $S$ stands for the fluid surface
$\tilde{z}=\tilde{\eta}(\tilde{r})$, $V$ is the
volume of a region bounded from above by the conic
surface $\tilde{z}=-s_0\tilde{r}$ and from below by the $S$
surface, and $\partial_n$ denotes the derivative along the normal to $S$. 
The integral on the right-hand side of this expression is
equal to the fluid velocity flux through the surface.
Since the function $\tilde{\Phi}$ 
is harmonic, the vector-field flux $\nabla\tilde{\Phi}$ 
through any closed surface is zero. This fact allows
the flux through the surface $S$ to be determined using
the asymptotic form of velocity potential at $R\to\infty$. 
Taking into account that the fluid flows into a solid
angle $2\pi(1+\cos\theta_0)$ at infinity, one has from Eq. (23)
$$
\int_S\partial_n\tilde{\Phi}\,dS=2\pi s(1+\cos\theta_0),
$$
and, hence,  $V=\pi s(1+\cos\theta_0)$. Notice that the volume
of a region bounded by the conic surface $\tilde{z}=-s_0\tilde{r}$ and 
the plane $\tilde z=-h$ (a circular right cone of height $h$) is
equal to $V$ at  
$$
h=h(s)=\left[3s{s_0}^2(1+\cos\theta_0)\right]^{1/3}.
$$
Clearly, if the volume  $V$ is fixed and the conditions
$\tilde{\eta}(\tilde r)+s_0\tilde{r}<0$ and 
$\tilde{\eta}_{\tilde r}(\tilde r)\leq 0$ are fulfilled for any 
$\tilde r$, the quantity  $|\tilde\eta(0)|$ 
cannot exceed the cone height. That is, the inequality
$$
|\tilde\eta(0)|\leq h(s),
$$
connecting the characteristic spatial scale at small $R$
with the asymptotic parameter $s$ is satisfied. Since the
most probable value of $h(s)$ corresponds to the maximum
allowable value $s_0$ of the $s$ constant, the following
estimate is also valid:
$$
|\tilde\eta(0)|\leq h(s_0)=s_0(3+3\cos\theta_0)^{1/3},
$$
which does not involves the free parameter $s$.

Let us return to the question of the applicability of
approximation (3)--(9) to the initial equations of
motion. As was pointed out above, condition (2) can be
used instead of Eq. (1) only if the external electric field
is weaker than the intrinsic cusp field. This implies that
the inequality $\varphi_r^2+\varphi_z^2\gg 1$ must be fulfilled.  
After the transition to the self-similar variables, it is recast as
$$
\tilde{\varphi}_{\tilde{r}}^2+\tilde{\varphi}_{\tilde{z}}^2\gg \tau^{2/3}.
$$
It is clear that for small $\tau$ (i.e., immediately before the
collapse) this condition is fulfilled near the singularity in
a natural way. Because one can write $|\nabla\varphi|\sim(r^2+z^2)^{-1/4}$
at small $\tau$, one has $|\nabla\varphi|\gg 1$ 
in a rather close vicinity of the singularity. 
In this case, $R_0$ and $\tau_0$ values exist for
which model (3)--(9) with $0\leq r^2+z^2<R_0^2$ and $0\leq\tau<\tau_0$
adequately describes the strongly nonlinear stages of
electrohydrodynamic instability development for the
free surface of a conducting fluid in an external electric
field. At $r^2+z^2>R_0^2$, the role of nonlinear processes is
rather insignificant; the condition for field uniformity at
infinity (1) should be used together with the corresponding
conditions in the limit $r\to\infty$, and, in particular,
with the condition for spatial localization of surface
perturbation: $\eta\to 0$ at $r\to\infty$. In this region,
the evolution of fluid surface is described by perturbation
theory for small surface slope; this procedure was
implemented, e.g., in \cite{10, 11}.

In conclusion, let us discuss the possibility to form
stronger singularities-cuspidal points. It is known
\cite{12} that the field mainly increases as $r^{-1}$
upon
approaching the apex of a thin point and, hence, the
electrostatic pressure $P_E$ increases as $r^{-2}$. 
Since the surface pressure changes as $P_S\sim r^{-1}$, $P_E\gg P_S$ 
near the singularity,
and the capillary effects can be ignored. 
It was shown in \cite{10} that in the absence of surface tension
weak root singularities $\eta\sim r^{3/2}$ form, for which the curvature
is infinite, while the surface itself remains
smooth. Therefore, when assuming that the cuspidal
points may appear, one arrives at a contradiction. This
gives grounds to assume that conic singularities are
precisely those which are the generic singular solutions
of the electrohydrodynamic equations, so that the
behavior of a charged liquid-metal surface with cusps is
described by self-similar solutions (10)--(21).

Note also that the results of this work can be
extended to dielectric fluids, in which the conic cusps
with angles depending on the dielectric constant can
form in an electric field \cite{13}. In addition, the approach
developed in this work can be applied to the description
of the evolution of dimples sharpening in a finite time
at a liquid helium surface (see, e.g., experimental work
\cite{14}). In my preceding work \cite{15} devoted to the construction
of exact analytic solutions to the equations of
motion for liquid helium in the presence of weak capillary
effects, I proved that cuspidal points $\eta\sim |x|^{2/3}$
appear at the surface in planar geometry. The question
of the singularity type for axial symmetry has not been
considered so far.

\medskip
The author are grateful to A.B.~Borisov and 
E.A.~Kuznetsov for interest in this work, and
V.G.~Suvorov for stimulating discussions.
The work was supported by the Russian Fund for Fundamental Research
(Project No.~00-02-17428) and, partly, by the INTAS Fund (Project
No.~99-1068).

\end {document}